\documentclass[twocolumn,aps]{revtex4}
\usepackage[english]{babel}
\usepackage{amssymb}
\usepackage{latexsym}
\usepackage{epsfig}
\usepackage{color}
\usepackage{float}
\usepackage{graphicx}
\usepackage{epstopdf}
\usepackage{natbib}
\usepackage{amsmath}
\usepackage[hidelinks=true]{hyperref}
\hypersetup{
  colorlinks   = true, %Colours links instead of  boxes
  urlcolor     = red, %Colour for external hyperlinks
  linkcolor    = red, %Colour of internal links
  citecolor   = blue %Colour of citations
}

\begin{document}

\title{Tsallis holographic dark energy in Fractal Universe}
\author{S. Ghaffari$^1$\footnote{sh.ghaffari@riaam.ac.ir}, E. Sadri$^2$\footnote{ehsan@sadri.id.ir}, A. H. Ziaie$^1$\footnote{ah.ziaie@riaam.ac.ir}}
\affiliation{$^1$ Research Institute for Astronomy and Astrophysics of Maragha (RIAAM), Maragha 55134-441, Iran}
\affiliation{$^2$ Azad University Central Tehran Branch, Tehran 34353-17117, Iran}

\begin{abstract}
We study the cosmological consequences of interacting Tsallis holographic dark energy model
in the framework of the fractal universe, in which, the Hubble radius is considered as the IR cut-off.
We drive the equation of state (EoS) parameter, deceleration parameter and the evolution equation for
the Tsallis holographic dark energy density parameter.
Our study shows that this model can describe the current accelerating Universe in both
noninteracting and interacting scenarios, and also a transition occurs
from the deceleration phase to the accelerated phase, at the late time.
Finally, we check the compatibility of free parameters of the model with the latest observational results
by using the Pantheon supernovae data, eBOSS, 6df, BOSS DR12, CMB Planck 2015, Gamma-Ray Burst.
\end{abstract}

\maketitle

\section{Introduction}
Since 1998 various observational data obviously suggest a mysterious type of energy with negative pressure, namely dark energy, is needed to describe the current accelerated expansion of the Universe~\citep{riess,garnavich,perlmutter,riess2,tegmark,spergel,Calabrese,ywang,mzhao}. Many efforts have been performed to study the dark energy while its nature still remains unknown~\citep{bamba,lili,wang,Peebles,padman,Copeland,Mli}. Energy of quantum fields in vacuum, bounded by the holographic hypothesis, is a pioneering candidate to model the dark energy~\cite{HDE, li,stab,RevH}. Due to the long range nature of gravity, it has been proposed that the generalized entropy formalism can be used to study the gravitational and cosmological phenomena~\cite{non2,CI,5,Tsallis,non4,non5,non6,non7,non13,non19,non20,non21,GSM,GSM1,GSM2,GSM3}. In this regard, a new holographic dark energy model has been proposed by using the holographic hypothesis and the Tsallis entropy, named Tsallis holographic dark energy (THDE)~\cite{Tavayef,thdb,THDE1,THDE3,THDE4,THDE5,mpla,sara}. Much attempts including other generalized entropies can be followed in~\cite{smm,epjcr,hooman}.\par THDE is proposed as~\cite{Tavayef}
\begin{eqnarray}\label{THDED}
\rho_{D}=\frac{3}{8\pi}BL^{2\delta-4},
\end{eqnarray}
in which $L$ denotes the IR cutoff, $\delta$ is a free parameter and $B$ is an unknown constant as usual. It is worthy to mention that observations allow a mutual interaction between the dark sectors of the cosmos including dark energy and dark matter which can even solve the coincidence problem~\cite{RevH,Int}. In this regard, it seems that for a cosmos filled by a dark energy fluid with density $\rho_D$ and a dark matter component of $\rho_m$, the mutual interaction $Q =3Hb^2(\rho_D + \rho_m)$, where $b^2$ is a coupling constant, is allowed which can even solve the coincidence problem~\cite{Int1}.

On the other hand, a fractal structure for the spacetime has been proposed~\cite{Frac1,Frac2,Frac3}, attracted a big deal of study~\cite{Clausius,Weinberg,Gallavotti,Gastmans,Aida,Christensen}. Also, it seems that the holographic dark energy models lead to interesting outcomes in the fractal cosmology~\cite{Sadri,Bolotin,Saavedra,Salti,Karami}. The action of Einstein gravity in fractal space–time is given by \citep{Frac1}
\begin{eqnarray}\label{action}
{\mathcal S}=\int d^4x\sqrt{-g}\Big(\frac{R-\omega\partial_\mu\nu\partial^\mu\nu}{2\kappa^2}+{\mathcal L}_m\Big),
\end{eqnarray}
\noindent in which $\kappa^2=8\pi G $, is the Einstein's gravitational constant, $ g $, $ R $ and $ {\mathcal L}_m $ are the Ricci curvature scalar,
determinant of the metric tensor $ g_{\mu\nu} $ and matter part of total Lagrangian density, respectively. The first Friedmann equation corresponding to the action~(\ref{action}) is then obtained as \citep{Frac1}
\begin{eqnarray}\label{Fried1}
H^2+H\frac{\dot{\nu}}{\nu}-\frac{\omega}{6}\dot{\nu}^2=\frac{1}{3M_p^2}(\rho_m+\rho_D),
\end{eqnarray}
\noindent where an overdot means derivative with respect to time, $ H=\dot{a}/a $ is the Hubble parameter, $ M_p^{-2}= 8\pi G $ is the reduced Planck mass, $\rho $ is the total energy density of the fluid filling the cosmos. Additionally, $ \omega $ is known as the fractal parameter, and $\nu$ determines the fractal function chosen in a power-law form as $ \nu=a^{-\beta} $ with $ \beta $ being a positive constant \citep{Frac1,Frac2,Frac3}.
%In addition, $ \nu $ determines the fractal function and $ \omega $ is known as the fractal parameter. This term results in a divergent behavior in the limit of %$t\rightarrow0$ as $\beta>0$. Therefore, presuming a suitable approximate relation between time and the scale factor in the total expansion of the universe ($a(t)\approx %t$) we can neglect the divergency and rewrite it as
%\begin{eqnarray}\label{nu}
%\nu=a^{-\beta}.
%\end{eqnarray}

\indent Motivated by the above arguments, here, we are interested in studying the evolution of a fractal Universe filled by a dark matter and THDE for two cases including $i)$ whenever there is no interaction between cosmic sectors, and $ii)$ the interaction is present. To reach this aim, we manage the article as follows: In the next section we present some general features of our model. A cosmography survey will be addressed in Sec.~(\textmd{III}). The statefinder diagnosis pair $ s-r $ constructs the backbone of our analysis in Sec. (\textmd{IV}). In Sec.~(\textmd{V}), using the recent observational data sets (SN Ia + BAO + CMB + OHD),
we fit the relevant free parameters, by employing the Markov-Chain-Monte Carlo (MCMC) method. The last section is devoted to a summary and concluding remarks.
%%%%%%%%%%%%%%%%%%%%%%%%%%%%%%%%%%%%%%%%%%%%%%%%%%%%%%%%%%%%%%%%%%%%%
%%%%%%%%%%%%%%%%%%%%%%%%%%%%%%%%%%%%%%%%%%%%%%%%%%%%%%%%%%%%%%%%%%%%%
\section{Fractal Cosmology}
\noindent We assume that there is mutual interaction between the dark sectors of the fractal Universe, then the conservation equations for dark matter and dark energy
are given by
\begin{eqnarray}\label{DMCons}
\dot{\rho}_m+(3-\beta)H\rho_m=Q,
\end{eqnarray}
\begin{eqnarray}\label{DECons}
\dot{\rho}_D+(1+\omega_D)(3-\beta)H\rho_D=-Q,
\end{eqnarray}
\noindent where $ w_D={p_D}/{\rho_D}$ is the EoS parameter of the THDE and $ \rho_m $ and $ \rho_D $ are the energy densities of dark matter and
dark energy, respectively. The quantity $Q$ denotes interaction between dark sectors. In fact, there are different choices for $Q$ term in order to study the dynamics of interacting DE models. In the recent work, the different phenomenological linear and non-linear interaction cases
in the framework of the holographic Ricci dark energy
model have been investigated and the results show that the linear interaction
$Q =3Hb\rho_D$ is the best case among the others~\citep{intcomp}.
Accordingly, in the herein model, we take $Q =3Hb^2\rho_D$ as the interaction term,
in which $ b^2 $ is a coupling constant.

\begin{figure}[!]
\begin{center}
\includegraphics[width=8cm]{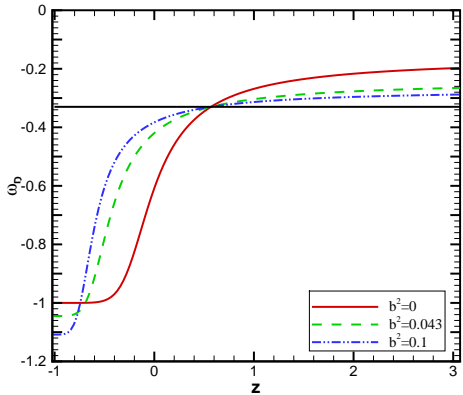}
\caption{The evolution of the EoS parameter $ \omega_D $ versus
redshift parameter $ z $. According to the best fitted values listed in Table \ref{bfs}, we have taken
$ \beta=0\cdot123 $, $\delta=1\cdot36$, $ \omega=0\cdot263 $ and some values of $b^2$.}\label{figw1}
\end{center}
\end{figure}

\noindent Defining the critical density as $\rho_{cr}={3H^2}/{8\pi}$ (we set the units so that $G=c=\hbar=1$) and the
density parameters
\begin{eqnarray}\label{Omega}
&&\Omega_m=\frac{\rho_m}{\rho_{cr}},\nonumber\\
&&\Omega_D=\frac{\rho_D}{\rho_{cr}},\nonumber\\
&&\gamma=-\beta-\frac{\beta^2\omega}{6}(1+z)^{2\beta},
\end{eqnarray}
along with using Eq.~(\ref{Fried1}), we get
\begin{eqnarray}\label{Fried2}
\Omega_m+\Omega_D=1+\gamma,
\end{eqnarray}
where $1+z=a^{-1}$ being the redshift.
\noindent Here we consider the Hubble radius as the IR cutoff i.e., $L=H^{-1}$, then
the energy density of THDE (\ref{THDED}) reads
\begin{equation}\label{rho1}
\rho_D=\frac{3}{8\pi}BH^{4-2\delta}.
\end{equation}
Taking the time derivative of Eq.~(\ref{rho1}) and combining the result with
Eqs.~(\ref{Fried2}) and~(\ref{DECons}) yields
\begin{eqnarray}\label{EoS1}
\omega_D=-1-\frac{3b^2}{(3-\beta)}+\frac{2\delta-4}{3-\beta}\frac{\dot{H}}{H^2}.
\end{eqnarray}
\noindent Next, we take the time derivative of Friedmann equation (\ref{Fried1}),
along with using Eqs.~(\ref{rho1}),~(\ref{Fried2}) and~(\ref{DMCons}), we get
\begin{equation}\label{dotH}
\frac{\dot{H}}{H^2}=\\
\frac{(3b^2-\beta+3)\Omega_D+(\beta-3)(1+\gamma)-\frac{\beta^3\omega}{3}(1+z)^{2\beta}}
{(2\delta-4)\Omega_D+2(1-\beta)-\frac{\beta^2\omega}{3}(1+z)^{2\beta}}.
\end{equation}
Moreover, the time derivative of Eq.~(\ref{Omega}) gives
\begin{eqnarray}\label{dotOmega1}
\dot{\Omega}_D=2(1-\delta)\Omega_D\frac{\dot{H}}{H^2}.
\end{eqnarray}
Inserting Eq. (\ref{dotH}) into Eq.~(\ref{dotOmega1}), the evolution of dimensionless THDE density parameter can be written as
%the equation of motion for the dimensionless GHDE density can be written as
%Using Eqs. (\ref{THDED} )and (\ref{dotH}), we can obtain
%the evolution of dimensionless THDE density as
\begin{figure}[!]
\begin{center}
    \includegraphics[width=8cm]{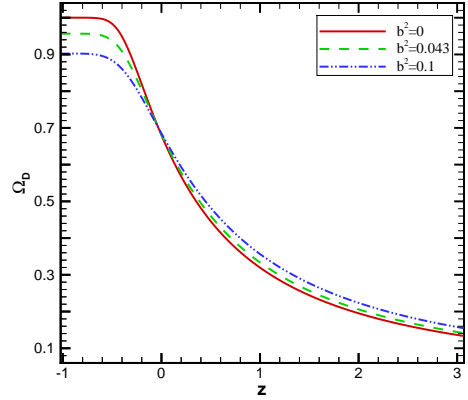}
    \caption{The evolution of density of dark energy versus $z$. According to the best fitted values listed in Table \ref{bfs}, we have taken
$H_0=68.87$, $\Omega_D=0.687$, $ \beta=0\cdot123 $, $\delta=1\cdot36$, $ \omega=0\cdot263 $ and some values of $b^2$.}\label{figomg}
\end{center}
\end{figure}
\begin{eqnarray}\label{dotOmega2}
&&{\Omega}^\prime_D=\frac{\dot{\Omega}_D}{H}=\Omega_D(1-\delta)\times\\&&
\frac{\Big((3b^2-\beta+3)\Omega_D+(\beta-3)(1+\gamma)-\frac{\beta^3\omega}{3}(1+z)^{2\beta}\Big)}
{(\delta-2)\Omega_D-\frac{\beta^2\omega}{6}(1+z)^{2\beta}-\beta+1},\nonumber
\end{eqnarray}
\noindent where the prime denotes derivative with respect to $ x=\ln a $.
In the limiting case $ \beta\rightarrow0 $,
The evolution of $ \Omega_D $ against the redshift $ z $, according to the best
fitted values of free parameters (Table \ref{bfs}), is plotted in Fig.~\ref{figomg}.
It can be easily seen that at early Universe ($ z\rightarrow\infty$)
we have $ \Omega_D\rightarrow 0 $, while at the late time $ (z\rightarrow 0) $
we get $ \Omega_D\rightarrow 1 $ .\\
Combining Eqs.~(\ref{rho1}),~(\ref{DECons}) and~(\ref{dotH}), one can obtain the EoS parameter as
\begin{eqnarray}\label{w1}
&&\omega_D=-1-\frac{3b^2}{(3-\beta)}+\\&&
\frac{(\delta-2)\Big((3b^2-\beta+3)\Omega_D+(\beta-3)(1+\gamma)-\frac{\beta^3\omega}{3}(1+z)^{2\beta}\Big)}
{(3-\beta)\Big((\delta-2)\Omega_D+1-\beta-\frac{\beta^2\omega}{6}(1+z)^{2\beta}\Big)}.\nonumber
\end{eqnarray}
One can easily see that for $ \beta\rightarrow0 $, where the effects of the fractal universe are negligible,
the EoS parameter of THDE in standard cosmology is recovered~\cite{Tavayef}.
The evolution of $ \omega_D(z) $ has been plotted in Fig.~\ref{figw1} for both interacting
and non-interacting cases. From this figure, it is clear that the model can describe the current accelerated
universe even in the absence of an interaction between two dark components,
and the transition redshift from the deceleration
phase to an accelerated phase occurs within the interval~$0.5<z<0.9 $,
which is in agreement with the recent observations~\cite{Daly,Komatsu,Salvatelli}.
We also have, in the non-interacting case (i.e., $ b^2=0 $), $ \omega_D(z\rightarrow0)\rightarrow -1 $
which means that THDE model in the fractal universe emulates the cosmological constant
while the interacting THDE model can cross the phantom line $  (\omega_D<-1)  $ at the late time.

\section{Cosmography}
The scale factor as the important component of studying the kinematics of the Universe is
accountable for the dependency of spatial separation in cosmological criterion.
Expanding the scale factor by the use of Taylor series in the proximity of the present time,
one can write
\begin{equation}\label{new_H}
a(t)=\sum\limits_{i=1}^\infty \frac{d^ia}{k!\,dt^i}(t-t_0)^i+1.
\end{equation}
According to the above definition of scale factor, three terms of cosmography series can be expressed as follows
\begin{equation}\label{H}
H(t)=\frac{1}{a}\frac{da}{dt},
\end{equation}
\begin{equation}\label{q}
q(t)=-\frac{1}{aH^2}\frac{d^2a}{dt^2}=-1-\frac{\dot{H}}{H^2},
\end{equation}
\begin{equation}\label{j}
j(t)=-\frac{1}{aH^3}\frac{d^3a}{dt^3}=q+2q^2+\frac{\dot{q}}{H}.
\end{equation}
Extending Eq. (\ref{new_H}) to higher order terms, one can reach the other parameters such as
Snap parameter ($s$) for $i=4$, which is helpful to study the deviation of the evolution of
the Universe from the $\Lambda$CDM. In the present work, we restrict the derivatives
to $i=1,2,3$, namely as the Hubble parameter, the deceleration parameter and the jerk parameter,
respectively. The deceleration parameter, i.e., the second derivative of the scale factor
with respect to cosmic time, makes it possible to check the behavior of expansion of the Universe.
Additionally, the transition redshift $z_t$ (at which $q(z_t)=0$) can be studied when the Universe
switches over from the decelerating to accelerating era. The cosmic jerk parameter $j$ as
the third derivative of the scale factor function provides a comparison between different models
of dark energy and $\Lambda$CDM ($j_0=1$). Hence, compared with the negative values of
the deceleration parameter, which indicates an accelerating Universe, the positive values of
the jerk parameter show an accelerating rate of the expansion. \\

\begin{figure}[!]
\begin{center}
\includegraphics[width=10cm]{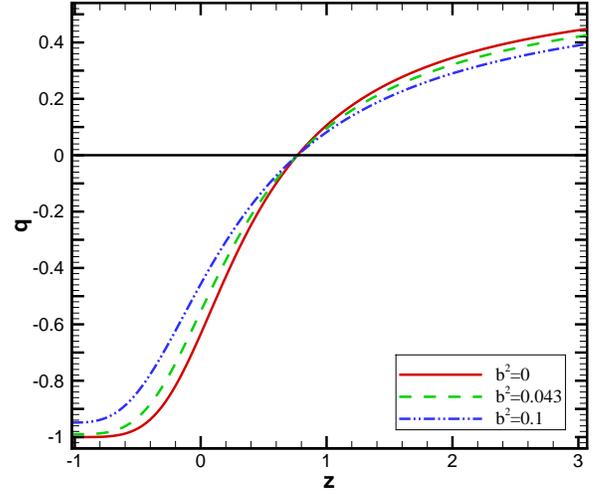}
\caption{The evolution of deceleration parameter $ q $ versus
redshift parameter $ z $. According to the best fitted values listed in Table \ref{bfs}, we have taken
$ \beta=0\cdot123 $, $\delta=1\cdot36$, $ \omega=0\cdot263 $ and some values of $b^2$.}\label{figq1}
\end{center}
\end{figure}

\noindent Using Eqs. (\ref{q}) and (\ref{dotH}), the deceleration parameter can be expressed as
\begin{eqnarray}\label{q1}
&&q=-1\\&&\nonumber
-\frac{(3b^2-\beta+3)\Omega_D+(\beta-3)(1+\gamma)-\frac{\beta^3\omega}{3}(1+z)^{2\beta}}
{(2\delta-4)\Omega_D+2(1-\beta)-\frac{\beta^2\omega}{3}(1+z)^{2\beta}},
\end{eqnarray}
For the limiting case $ \beta\rightarrow0 $, the deceleration parameter of THDE in standard cosmology is recovered~\cite{Tavayef}.
According to the best values of the fitted parameters presented in Table \ref{bfs},
the value of the deceleration parameter at present time is $q_0\approx-0.55$,
for the interacting model which is consistent with the obtained
value of the deceleration parameter by Planck $(q_0=-0.55)$ \cite{Ade}
and demonstrates an accelerating expansion of the current Universe.
The behavior of the deceleration parameter $ q $ versus redshift $ z $ is plotted numerically in Fig. \ref{figq1}.
According to Fig. \ref{figq1} one may see that both models enter
the accelerating era at $z=0.7$ which is within the range $z=(0.4,0.8)$
obtained by the recent observational works~\citep{dec1,dec2,dec3,dec4,dec5}.\par
\begin{figure}[!]
\begin{center}
    \includegraphics[width=8cm]{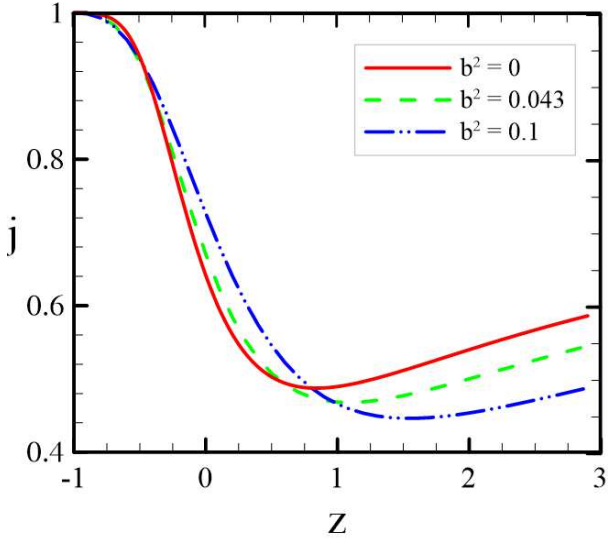}
    \caption{The evolution of the cosmic jerk parameter in terms of redshift. According to the best fitted values listed in Table \ref{bfs}, we have taken
  $ \beta=0\cdot123 $, $\delta=1\cdot36$, $ \omega=0\cdot263 $ and some values of $b^2$.}\label{figj}
\end{center}
\end{figure}

For the jerk parameter, by inserting Eq. (\ref{q1}) into Eq. (\ref{j})
and with help of Eqs. (\ref{dotH}) and (\ref{dotOmega2}) we have
\begin{eqnarray}\label{j1}
&&j=q+2q^2-\\&&\nonumber
\Bigg[\Big((3b^2-\beta+3)\Omega^\prime_D+(\beta-1)\beta^3\omega(1+z)^{2\beta}\Big)\\&&\nonumber \times\left((2\delta-4)\Omega_D-\frac{\beta^2\omega}{3}(1+z)^{2\beta}+2(1-\beta)\right)\\&&\nonumber
-\left((2\delta-4)\Omega^\prime_D+\frac{2\beta^3\omega}{3}(1+z)^{2\beta}\right)\\&&\nonumber
\times\Big((3b^2-\beta+3)\Omega_D+(\beta-3)(1+\gamma)-\frac{\beta^3\omega}{3}(1+z)^{2\beta}\Big)\Bigg]\\&&\nonumber\times \!\!\left(\!\!(2\delta-4)\Omega_D-\frac{\beta^2\omega}{3}(1+z)^{2\beta}+2(1-\beta)\right)^{-2}.\nonumber
\end{eqnarray}
As it is mentioned, in comparison with the deceleration parameter, the positive value of jerk parameter indicates an accelerated expansion of the Universe. The behavior of jerk parameter is plotted in Fig. (\ref{figj}). It is seen that this parameter stays positive and tends to unity at late time. According to the results of observational studies, the value of the cosmic jerk parameter has a weaker restriction compared to the deceleration parameter $-5<j_0<10$. In this work, using the observational data, we obtained the value of the cosmic jerk parameter for both models at the time of observation $j_0\approx0.69$. It is worthwhile to mention that the THDE model has a tendency toward 1 or the $\Lambda$CDM model at the late time.
%%%%%%%%%%%%%%%%%%%%%%%%%%%%%%%%%%%%%%%%%%%%%%%%%%%%
\section{ The statefinder pair}%%%%%%%%%%%%%%%%%%%%%%%%%%%%%%%%%%%%%%
Using the Hubble parameter for studying the evolution of cosmic expansion (Eq. \ref{dotH}) and the deceleration parameter for surveying the rate of acceleration and deceleration of cosmic expansion (Eq. \ref{q1}), we cannot clearly discern various dark energy models using these two parameters once $H > 0$ or $q < 0$. Calling for more accurate calculations regarding this issue and resulting from the improvement of observational data during the recent two decades, a new geometrical diagnostic pair for tracking the dark energy models has been proposed \citep{{sr1},{sr2}}. This new pair letting us to specify the characteristics of dark energy is called statefinder pair $(r,s)$
\begin{eqnarray}\label{rs}
&&r=\frac{\dddot{a}}{aH^3}=1+\frac{\ddot{H}}{H^3}+3\frac{\dot{H}}{H^2},\nonumber\\
&&s=\frac{r-1}{3\left(q-\frac{1}{2}\right)}.
\end{eqnarray}
In order to investigate the statefinder for NHDE in the framework of fractal cosmology, we must obtain $\frac{\ddot{H}}{H^3}$. Consequently we can calculate ($s$). Taking the time derivative of both sides of Eq.~(\ref{dotH}) we get
\begin{eqnarray}\label{ddhz}
&&\frac{\ddot{H}}{H^3}=\Bigg[\Big((3b^2-\beta+3)\Omega^\prime_D+(\beta-1)\beta^3\omega(1+z)^{2\beta}\Big)\nonumber\\&&\nonumber
\times\left((2\delta-4)\Omega_D+2(1-\beta)-(\beta^2\frac{\omega}{3})(1+z)^{2\beta}\right)\\&&\nonumber
-\left((2\delta-4)\Omega^\prime_D+\frac{2\beta^3\omega}{3}(1+z)^{2\beta}\right)\\&&\nonumber
\times\Big((3b^2-\beta+3)\Omega_D+(\beta-3)(1+\gamma)-\frac{\beta^3\omega}{3}(1+z)^{2\beta}\Big)\Bigg]\\&&\nonumber
\times\Big((2\delta-4)\Omega_D +2(1-\beta)
-(\beta^2\frac{\omega}{3})(1+z)^{2\beta}\Big)^{-2}\\&&+2\left(\frac{\dot{H}}{H^2}\right)^2,
\end{eqnarray}
in which $\Omega^\prime_D=\frac{\dot{\Omega}_D}{H}$. It can be seen that as the Universe expands the value of parameter $r$ increases for both interacting and non-interacting models and stays smaller than unity. The parameter $s$ during the whole evolution stays in positive region for both interacting and non-interacting models. The fixed point $ \left(r,s\right) = \left(1,0\right)$ represents the $\Lambda$CDM scenario. Tracing each case demonstrates that interacting and non-interacting models have the quintessence behavior $\left(s>0,r<1\right)$. Furthermore, the trajectories of both models meet the fixed point $\left(1,0\right)$ indicating the evolution from quintessence to phantom-like behavior as the Universe expands. The present value of both interacting and non-interacting models roughly coincide each other and have the identical distance from the $\Lambda$CDM fixed point. The results of statefinder for the THDE model is an affirmation on the results of equation of state (Eq.~\ref{EoS1}).
\begin{figure}[H]
\centering
\begin{tabular}{ccc}
\hspace*{-0.1in}
\includegraphics[width=8cm]{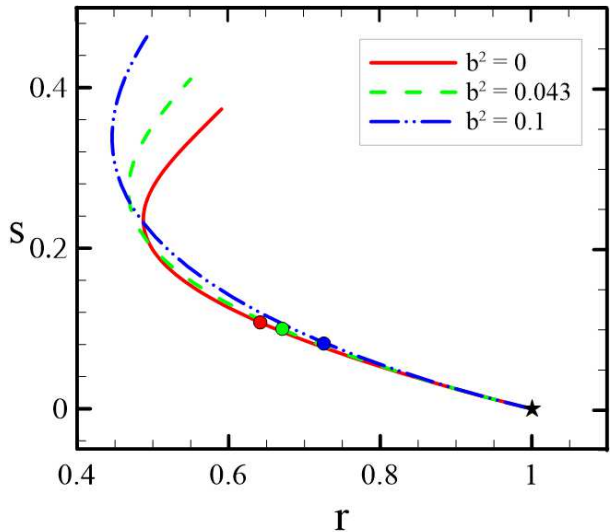}
\end{tabular}
\caption{\small The evolution of parameter $s$ in terms of parameter $r$.
According to the best fitted values listed in Table \ref{bfs}, the different parameter values
 $ \beta=0\cdot123 $, $\delta=1\cdot36$, $ \omega=0\cdot263 $ are adopted.
The star symbol denotes the $\Lambda$CDM model and dot symbols represent the present value
of each model for different values of coupling constant $b^2$.} \label{sr}
\end{figure}\label{rs}
%%%%%%%%%%%%%%%%%%%%%%%%%%%%%%%%%%%%%%%%%%%%%%%%%%%%%%%%%%%%%%%%%%%%%%%%%%%%%%%%%%
\section{DATA SETS}
In order to find the best values of the current model we combine the latest observational data including SNIa, BAO and CMB. For this purpose, we use the public codes EMCEE \citep{FM} for performing the MCMC method and GetDist Python package~\footnote{https://getdist.readthedocs.io} to plot and analyze the contours. This method also provides reliable error estimates on the measured variables. For all analysis we chose 550 iterations and 400 walkers to make a chain with $22\times10^4$ points.\\
\subsection{TYPE IA SUPERNOVAE}
 As a large-scale investigator of the cosmic expansion, observations of type Ia supernovae play an important role to map the expansion history of the Universe. In this work, we use 1048 data points of the recent proposed Pantheon Supernovae project \citep{Scolnic} comprising the redshift range $0.01<z<2.3$. We use the systematic covariance $C_{sys}$ as
\begin{equation}\label{SNsys}
C_{ij,sys}=\sum_{n=1}^{i}\left(\frac{\partial \mu_i}{\partial S_n}\right)\left(\frac{\partial \mu_j}{\partial S_n}\right)\left(\sigma_{S_k}\right),
\end{equation}
in which the summation is performed over the $n$ systematic components $S_n$ and the related magnitude of error $\sigma_{S_n}$. According to $\triangle\mu=\mu_{data}-M-\mu_{obs}$ where $M$ is a nuisance parameter one may write the $\chi^2$ relation for Pantheon SNIa data as
\begin{equation}\label{SNchi2}
\chi^2_{Pantheon}=\triangle\mu^T\cdot C_{Pantheon}^{-1}\cdot\triangle\mu.
\end{equation}
It should be noted that the $C_{Pantheon}$ is the summation of the systematic covariance and statistical matrix $D_{stat}$ having a diagonal component. The complete version of full (1048 data points) and binned (40 data points) Pantheon supernova data can be found in the online source~\footnote{https://archive.stsci.edu/prepds/ps1cosmo/index.html}. The binned data is a good approximation of the full data but after implementing the MCMC method the value of $\chi_{dof}$ for binned data is bigger than the full data and shows the lower accuracy compared to the full range of data.\\
%%%%%%%%%%%%%%%%%%%%%%%%%%%%%%%%%%%%%%%%%%%%%%%%
%%%%%%%%%%%%%%%%%%%%%%%%%%%%%%%%%%%%%%%%%%%%%%%%
\subsection{GAMMA-RAY BURST} %https://arxiv.org/pdf/1103.0378.pdf
Similar to the SNIA data one can constrain the free parameters values using GRB  by fitting the distance modulus $\mu(z)$. In this work we use 109 data of Gamma-Ray Burst in the redshift range $0.3<z<8.1$\citep{grbref} embracing the 50 low-z GRBs ($z<1.4$) and high-z GRBs ($z>1.4$). The 70 data of GRBs are obtained from \citep{amat1}, 25 GRBs are taken from\citep{amat2} and the remains 14 GRBs data points are extracted from \citep{amat3}. The $\chi^2$ for GRB is given by
\begin{equation}\label{grbfunc}
\chi^2_{GRB}=\sum_{i=1}^{109} \frac{\left[\mu_{obs}(z_i)-\mu_{th}(z_i)\right]^2}{\sigma^2(z_i)},
\end{equation}
in which the theoretical distance modulus $\mu_{th}(z_i)$ can be defined as
\begin{equation}\label{grbfunc}
\mu_{th}(z_i)=5log_{10}D_l(z_i)+\mu_0.
\end{equation}
where $\mu_0=42.38-5log_{10}h$ and $h=H_0/100$ with unit of km/s/Mpc and $H_0$ is the value of Hubble parameter (Hubble constant) at the present time or time of observation.
\subsection{BARYON ACOUSTIC OSCILLATIONS}
The baryon acoustic oscillations (BAO) are the large-scale impression of oscillations in the early time plasma and are strong standard ruler to measure the angular diameter distance. In this paper, we combine the extended Baryon Oscillation Spectroscopic Survey (eBOSS) quasar clustering at $z=1.52$ \citep{Beutler}, isotropic BAO measurements of 6dF survey at an effective redshift ($z=0.106$) \citep{Ata} and the BOSS DR12 \cite{Alam} including six data points of Baryon Oscillations as the latest observational data for BAO. The $\chi^2_{BAO}$ of BOSS DR12 may be express as
\begin{equation}\label{BOSSDR12}
\chi^2_{BOSS~DR12}=X^tC_{BAO}^{-1}X,
\end{equation}
where $X$ for six data points is
\begin{equation}\label{XBAO}
X=\left(\begin{array}{c} \frac{D_M\left(0.38\right)r_{s,fid}}{r_s\left(z_d\right)}-1512.39\\
\frac{H\left(0.38\right)r_s\left(z_d\right)}{r_s\left(z_d\right)}-81.208\\
\frac{D_M\left(0.51\right)r_{s,fid}}{r_s\left(z_d\right)}-1975.22\\
\frac{H\left(0.51\right)r_s\left(z_d\right)}{r_s\left(z_d\right)}-90.9\\
\frac{D_M\left(0.61\right)r_{s,fid}}{r_s\left(z_d\right)}-2306.68\\
\frac{H\left(0.51\right)r_s\left(z_d\right)}{r_s\left(z_d\right)}-98.964\end{array}\right),
\end{equation}
and $r_{s,fid}=$147.78 Mpc is the sound horizon of the fiducial model, $D_M\left(z\right)=\left(1+z\right)D_A\left(z\right)$ is the comoving angular diameter distance and $Cov_{BAO}$  is the covariance matrix~\cite{Alam}. One can define the sound horizon at the decoupling time $r_s\left(z_d\right)$ as
\begin{equation}\label{BAO1}
r_s\left(z_d\right)=\int_{z_d}^{\infty} \frac{c_s\left(z\right)}{H\left(z\right)}dz,
\end{equation}
in which $c_s=1/\sqrt{3\left(1+R_b/\left(1+z\right)\right)}$ is the sound speed with
$R_b=31500\Omega_bh^2\left(2.726/2.7\right)^{-4}$.
The total $\chi^2$ for all baryonic acoustic oscillations data is
\begin{equation}\label{BAO}
\chi^2_{BAO}=\chi^2_{BOSS~DR12}+\chi^2_{6dF}+\chi^2_{eBOSS}.
\end{equation}
\subsection{COSMIC MICROWAVE BACKGROUND}
We study the Cosmic Microwave Background (CMB) to discover the expansion history of the Universe. For this, we use the data of Planck 2015 \cite{Ade}. The $\chi^2_{CMB}$ function can be defined as
\begin{equation}\label{CMB}
\chi^2_{CMB}=q_i-q^{data}_i Cov^{-1}_{CMB}\left(q_i,q_j\right),
\end{equation}
where $q_1=R\left(z_*\right)$, $q_2=l_A\left(z_*\right)$ and $q_3=\omega_b$ and $Cov_{CMB}$
is the covariance matrix \cite{Ade}. The data of Planck 2015 are
\begin{equation}\label{PLANCKDATA}
q^{data}_1=1.7382,~\\
q^{data}_2=301.63,~\\
q^{data}_3=0.02262.
\end{equation}
The acoustic scale $l_A$ is
\begin{equation}\label{lA}
l_A=\frac{3.14d_L\left(z_*\right)}{\left(1+z\right)r_s\left(z_*\right)},
\end{equation}
in which $r_s\left(z_*\right)$ is the comoving sound horizon at the drag epoch ($z_*$). The function of redshift at the drag epoch is \cite{whu}.
\begin{equation}\label{z_*}
z_*=1048\left[1+0.00124\left(\Omega_bh^2\right)^{-0.738}\right]\left[1+g_1\left(\Omega_mh^2\right)^{g_2}\right],
\end{equation}
where
\begin{equation}\label{g1 g2}
g_1=\frac{0.0783\left(\Omega_bh^2\right)^{-0.238}}{1+39.5\left(\Omega_bh^2\right)^{-0.763}}, ~~~g_2=\frac{0.560}{1+21.1\left(\Omega_bh^2\right)^{1.81}},
\end{equation}
The CMB shift parameter is \cite{ywangm}
\begin{equation}\label{R}
R=\sqrt{\Omega_{m_0}}\frac{H_0}{c}r_s\left(z_*\right).
\end{equation}
As an optimum way of constraining the wide range of dark energy models in this work we use the CMB data which does not contain the full Planck information.

The data for BAO and CMB could be found in the online source of latest version of MontePython \footnote[1]{http://baudren.github.io/montepython.html}.
Using minimized $\chi^2_{min}$, we may constrain and obtain the best-fit values of the free parameters
\begin{equation}\label{chi}
\chi_{min}^2=\chi_{SNIa}^2+\chi_{GRB}^2+\chi_{CMB}^2+\chi_{BAO}^2.
\end{equation}The best-fit values of $ H_0$, $\Omega_D$, $\delta$, $\omega$, $\beta$, $b$ and $M$ by consideration of the $1\sigma$ confidence level are shown in Table \ref{bfs}.

\begin{table}[H]%sidewaystable
%\footnotesize
\centering
\caption{The fitted values of cosmological parameters for the interacting and non-interacting THDE model in the framework of the fractal universe. For obtaining the values using MCMC method the Pantheon Supernovae data, BAO (BOSS DR12, 6df, eBOSS), CMB Planck 2015 and Gamma-Ray Burst data have been used. The $\chi_{dof}$ denotes the goodness of fit and could be obtained by the $\chi^2/(n-N)$ which $n$ is the number of data points (here 1168) and $N$ is the number of total free parameters (6 for non-interacting and 7 for interacting THDE model).}
\label{bfs}
\hspace*{-1.5em}
\footnotesize\addtolength{\tabcolsep}{3pt}
\begin{tabular}{cc}
\hline\hline
\multicolumn{2}{c}{ }                                             \\ [-0.4cm]
\multicolumn{2}{c}{ Tsallis Holographic Dark Energy Model}                                             \\ [-0.02cm] \hline
\multicolumn{1}{c}{Params}                 & \multicolumn{1}{c}{$NON-INTERACTING$}           \\ [-0.02cm]\hline
\multicolumn{1}{c}{}                              & \multicolumn{1}{l}{}                          \\  [-0.2cm]
\multicolumn{1}{c}{$H_0$ }   & \multicolumn{1}{c}{$68.783^{+0.961}_{-0.761}$}   \\ [0.15cm]
\multicolumn{1}{c}{$\Omega_D$ }             & \multicolumn{1}{c}{$0.687^{+0.024}_{-0.028}$} \\ [0.15cm]
\multicolumn{1}{c}{$\delta$}      & \multicolumn{1}{c}{$1.360^{+0.160}_{-0.191}$}     \\ [0.15cm]
\multicolumn{1}{c}{$\omega$}      & \multicolumn{1}{c}{$0.201 _ { - 0.029 } ^ { + 0.029 }$} \\ [0.15cm]
\multicolumn{1}{c}{$\beta$}      & \multicolumn{1}{c}{$0.123^{+0.059}_{-0.063}$}  \\ [0.15cm]
\multicolumn{1}{c}{$b^2$}      &  \multicolumn{1}{c}{$0.0423 _ { - 0.02 } ^ { + 0.02}$} \\  [0.15cm]
\multicolumn{1}{c}{$M$}      & \multicolumn{1}{c}{$-19.375^{+0.023}_{-0.019}$}    \\  [0.15cm]
\multicolumn{1}{c}{$\chi^2$}      & \multicolumn{1}{c}{$1104.6409$}    \\  [0.15cm]
\multicolumn{1}{c}{$X_{dof}$}      & \multicolumn{1}{c}{$0.9514$}  \\  [0.15cm] \hline\hline
\end{tabular}
\end{table}

\begin{figure}[htp]
\begin{center}
\includegraphics[width=9cm]{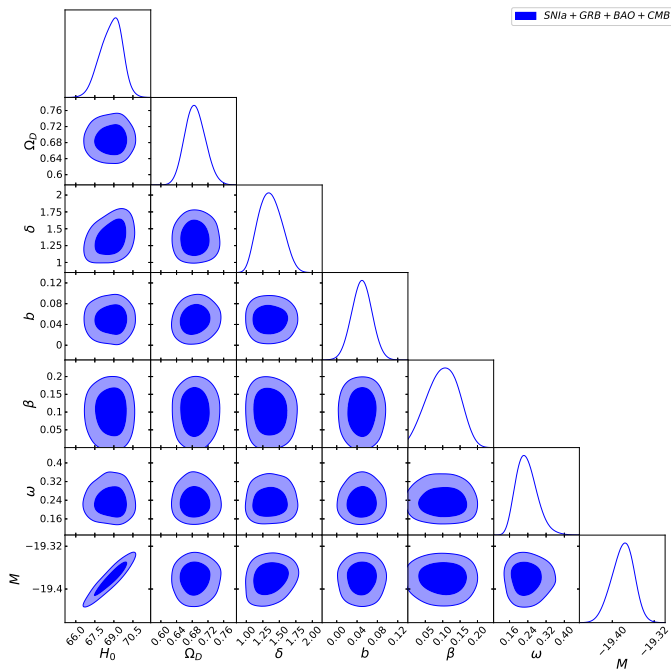}
\caption{The contour map of the interacting Tsallis holographic dark energy model in the framework of the fractal cosmology. In this figure $H_0$ is the Hubble parameter, $\Omega_D$ is the density of dark energy, $\delta$ denotes the free parameter related to the THDE model, $\beta$ is the free parameter of power law determining the fractal function, $b$ is the coupling constant from the interaction term, $\omega$ is the fractal parameter and $M$ is the nuisance parameter of Pantheon data. The best fitted value of these parameters are listed in the Table \ref{bfs}}\label{contint}
\end{center}
\end{figure}
The Hubble constant $H_0$ as an important quantity in cosmology for calculating the age and size of the Universe corresponds to the Hubble parameter at the present time or time of the observation. In this work, by the use of the latest observational data we observe that the obtained value for $H_0$ for the THDE model is in good agreement with latest obtained results for Hubble parameter $H_0=(67,70)$\citep{Ade,h1,h2,h3,h4}. We also found that the measured value for the density of dark energy has a good consistency with recent measured value for this component~\citep{Ade, h4, h5}.
For checking the success of the models on fitting data we may calculate the goodness of fit $\chi_{dof}$, we may check the success of the models on fitting data. The goodness of fit can be explained as $\chi^2/(n-N)$ in which $n=1168$ and $N$ represent the total number of data points and free parameters respectively. The THDE model with identical $\chi_{dof}$ is successful with the appropriate values (less than 1). These results prove that the THDE model in fractal cosmology has a good consistency with the latest observational data and the additional interaction term does not impose any problem to this issue.\\

\section{Summary and Concluding Remarks}
In this paper, we studied the Tsallis holographic dark energy
model with Hubble horizon as IR cutoff in the framework of the
flat fractal cosmology. We used full SNIa Pantheon data, the
extended Baryon Oscillation Spectroscopic Survey, quasar
clustering, 6df survey, BOSS DR12, Cosmic Microwave Background
(CMB) of Planck 2015, Gamma-Ray burst as the observational data
for constraining the free parameters of the model. For obtaining
the results we employed and modified the Cosmo Hammer (EMCEE)
Python package public code. We found that the deceleration
parameter for the THDE model demonstrates a Universe with
accelerating rate of expansion and it can be seen that the THDE
model enters the accelerating era within the redshift
$z=(0.6,0.8)$ which shows a good compatibility with recent studies
$0.5<z_t<1$.

The coupling constant $b$ which has been measured as a positive
and small value conveys the decay of the dark energy into dark
matter. The $r-s$ plane is plotted in Fig. \ref{sr} where we
observe that all trajectories for the THDE model meet the
$\Lambda$CDM fixed point $(r,s)=(1,0)$. The statefinder
trajectories indicate the quintessence behavior for both models
(where $s>0, r<1$) and also embracing the $\Lambda$CDM fixed point
denotes the transition from quintessence to phantom. This is
consistent with the results of the equation of state. Our results
demonstrated that the value of the Hubble constant is in range
$H_0=(68,70)$ having good agreement with the results of recent
works on observational data. We obtained the coupling constant as
a positive and small value indicating the possibilities of
decaying the dark energy into the dark matter.

\indent It should be noted that for the deep understanding of
behavior of THDE in the fractal Universe, specifically the
interacting model, more investigations should be done. Therefore,
for the future works, we would like to study the dynamical system
methods to figure out the status of the non-linear interactions in
the late time within the framework of the fractal universe.
Another point is to study the perturbation analysis compare to the
gravitational lenses and the Large Scale Structure.
%%%%%%%%%%%%%%%%%%%%%%%%%%%%%%%%%%%%%%%%%%%%%%%%%%%%%%
\section*{Acknowledgment}
The work of S. Ghaffari has been supported financially by Research
Institute for Astronomy \& Astrophysics of Maragha (RIAAM).
%%%%%%%%%%%%%%%%%%%%%%%%%%%%%%%%%%%%%%%%%%%%%%%%%%%%%%%
 \bibliographystyle{unsrtant}

\end{document}